\documentclass[12pt,a4paper]{JHEP3}
\preprint{MITP/13-002\\
ZU-TU 01/13\\
LPN13-007\\
MCnet-13-01
}

\pdfoutput=1

\usepackage{snapshot}
\usepackage{cite}
\usepackage{amssymb}
\usepackage{amsmath}
\usepackage{epsfig}
\usepackage{subfigure}
\usepackage{color}
\let\normalcolor\relax
\usepackage{graphicx}
\usepackage{inputenc}
\usepackage{xspace}
\usepackage{slashed}
\input{latin1.def}
\usepackage{axodraw4j}

\def\lsim{\mathrel{\raise.3ex\hbox{$<$\kern-.75em\lower1ex\hbox{$\sim$}}}}
\def\gsim{\mathrel{\raise.3ex\hbox{$>$\kern-.75em\lower1ex\hbox{$\sim$}}}}
\def\ifmath#1{\relax\ifmmode #1\else $#1$\fi}

\newcommand{\beq}{\begin{equation}}
\newcommand{\eeq}{\end{equation}}
\def\beqn{\begin{eqnarray}}
\def\eeqn{\end{eqnarray}}

\newcommand{\bea}{\begin{eqnarray}}
\newcommand{\eea}{\end{eqnarray}}

\title{\boldmath{Higgs Boson self-coupling measurements using ratios of
  cross sections}}

\author{Florian Goertz$^2$, Andreas Papaefstathiou$^3$, Li Lin Yang$^1$,
  Jos\'e Zurita$^4$\\
$^1$ School of Physics and State Key Laboratory of Nuclear Physics and Technology, Peking University, Beijing 100871, China.\\
$^2$ Institut f\"ur Theoretische Physik, ETH Z\"urich, 8093 Zurich,
Switzerland. \\
$^3$  Institut f\"ur Theoretische Physik, Universit\"at Z\"urich,
8057 Zurich, Switzerland.\\
$^4$ PRISMA Cluster of Excellence \& Mainz Institute for Theoretical Physics Johannes Gutenberg University, 55099 Mainz, Germany.
}

\abstract{We consider the ratio of cross sections of double-to-single Higgs boson
  production at the Large Hadron Collider at 14~TeV. Since both processes possess similar higher-order corrections, leading to a cancellation of uncertainties in the ratio, this observable is well-suited to constrain the trilinear Higgs boson self-coupling. We consider the scale variation, parton density function uncertainties and conservative estimates of
experimental uncertainties, applied to the viable decay channels, to
construct expected exclusion regions. We show that the trilinear self-coupling can be constrained to be
positive with a 600~fb$^{-1}$ LHC dataset at 95\%
confidence level. Moreover, we demonstrate that we expect to obtain a $\sim+30\%$
and $\sim-20\%$ uncertainty on the self-coupling at 3000~fb$^{-1}$ without
statistical fitting of differential distributions. The present article
outlines the most precise method of determination of the Higgs trilinear
coupling to date.}


\begin{document}
\section {Introduction}
One of the aims of the Large Hadron Collider (LHC) is to search for the agent of electroweak symmetry breaking (EWSB), which in
its minimal form is the Standard Model (SM) Higgs boson ($H$).
Recently, both the ATLAS and the CMS collaborations have observed a
new state with a mass of about 125~GeV, whose properties are in
substantial agreement with the SM Higgs boson~\cite{ATLAS_Higgs,
  CMS_Higgs, CMS-PAS-HIG-12-045, ATLAS:2012wma, ATLAS-CONF-2012-170}.
The quest for understanding the mechanism behind EWSB does not end
with the discovery of this particle. It is crucial to test the Higgs sector to its full extent, measuring the couplings of the Higgs
boson to gauge bosons and matter fields~\cite{Ellis:2012hz, Carmi:2012yp, Azatov:2012bz,
  Espinosa:2012ir, Giardino:2012ww, Azatov:2012rd, Klute:2012pu,
  Espinosa:2012vu, Carmi:2012zd, Chang:2012tb,
  Chang:2012gn, Low:2012rj, Corbett:2012dm, Giardino:2012dp,
  Montull:2012ik, Espinosa:2012im, Carmi:2012in, Banerjee:2012xc,
  Bonnet:2012nm, Plehn:2012iz, Djouadi:2012rh, Cacciapaglia:2012wb,
  Masso:2012eq, Gupta, Belanger}, and also to probe its
self-interactions~\cite{Baur:2002rb, Baur:2002qd, Dolan:2012rv,
  Papaefstathiou:2012qe, Baglio:2012np, Djouadi:1999rca}. After EWSB, the Higgs potential can be written as
\begin{equation}
V(H) = \frac{1}{2} M_H^2 H^2 + \lambda_{HHH} v H^3 + \frac{1}{4}
\lambda_{HHHH} H^4 \, .
\end{equation}
In the SM, $\lambda_{HHH}^{SM} =\lambda_{HHHH} ^{SM}= (M_H^2 / 2 v^2)
\approx 0.13$ for a Higgs mass of $M_H \simeq 125$~GeV and a vacuum
expectation value of $v\simeq 246 $~GeV. We can also define normalised couplings $\lambda \equiv
\lambda_{HHH} / \lambda_{HHH}^{SM}$ and $\tilde{\lambda} \equiv
\lambda_{HHHH} / \lambda_{HHHH}^{SM}$. 

A measurement of these two couplings is crucial to the reconstruction of the Higgs potential and will allow testing of the EWSB mechanism. Moreover, in many models beyond the SM, these couplings may deviate from the SM
values, and in that case they will provide relevant information about
the nature of the new physics model.

At the LHC, the quartic
coupling $\tilde{\lambda}$ may be probed via triple Higgs boson production.
However, its tiny cross section \cite{Plehn:2005nk} makes it very
difficult, if not impossible, to do so. On the other hand, the trilinear coupling
$\lambda$ can be measured in Higgs boson pair production, $pp \to
HH$, which may be discovered at a large luminosity phase of the LHC. 

The discovery potential for Higgs boson pair production at the LHC has
been studied in~\cite{Baur:2002qd, Baur:2003gp, Dolan:2012rv,
  Papaefstathiou:2012qe, Baglio:2012np}. In Refs.~\cite{Baur:2002qd, Baur:2003gp},
constraints were placed on $\lambda$ using statistical fits to the shape of the
visible mass distributions of the final decay products of the Higgs pairs, whereas Refs.~\cite{Dolan:2012rv,
  Papaefstathiou:2012qe} focused on the establishment of the Higgs
pair production process using improved techniques originating mainly
from developments in the understanding of boosted jet
substructure~\cite{Butterworth:2008iy, Abdesselam:2010pt}. In
Ref.~\cite{Baglio:2012np} the final state $b\bar{b}\gamma\gamma$ was
revisited as well as $b\bar{b}\tau^+\tau^-$ and $b\bar{b}W^+W^-$ (fully
leptonic), without making use of jet substructure techniques (although
boosted Higgs bosons were required). The present article concentrates on using the results
from the available phenomenological studies along with the best
available theoretical cross section calculations and conservative
estimates of the experimental uncertainties, to demonstrate the
possibility of constraining the trilinear Higgs self-coupling at the LHC. 

The article is organised in the following way: in
Section~\ref{sec:cross} we dissect the Higgs boson production cross
sections and in Section~\ref{sec:ratios} we examine the theoretical uncertainties
on the ratio of cross sections of double-to-single Higgs production. Then, in
Section~\ref{sec:constaining}, we present the expected constraints obtained at
integrated luminosities of 600~fb$^{-1}$ and 3000~fb$^{-1}$ for a
simplified model, as well as within the Standard Model. We conclude in Section~\ref{sec:conclusions}.

\section{Dissection of the cross sections}\label{sec:cross}
\begin{figure}[!htb]
    \includegraphics[width=0.5\linewidth]{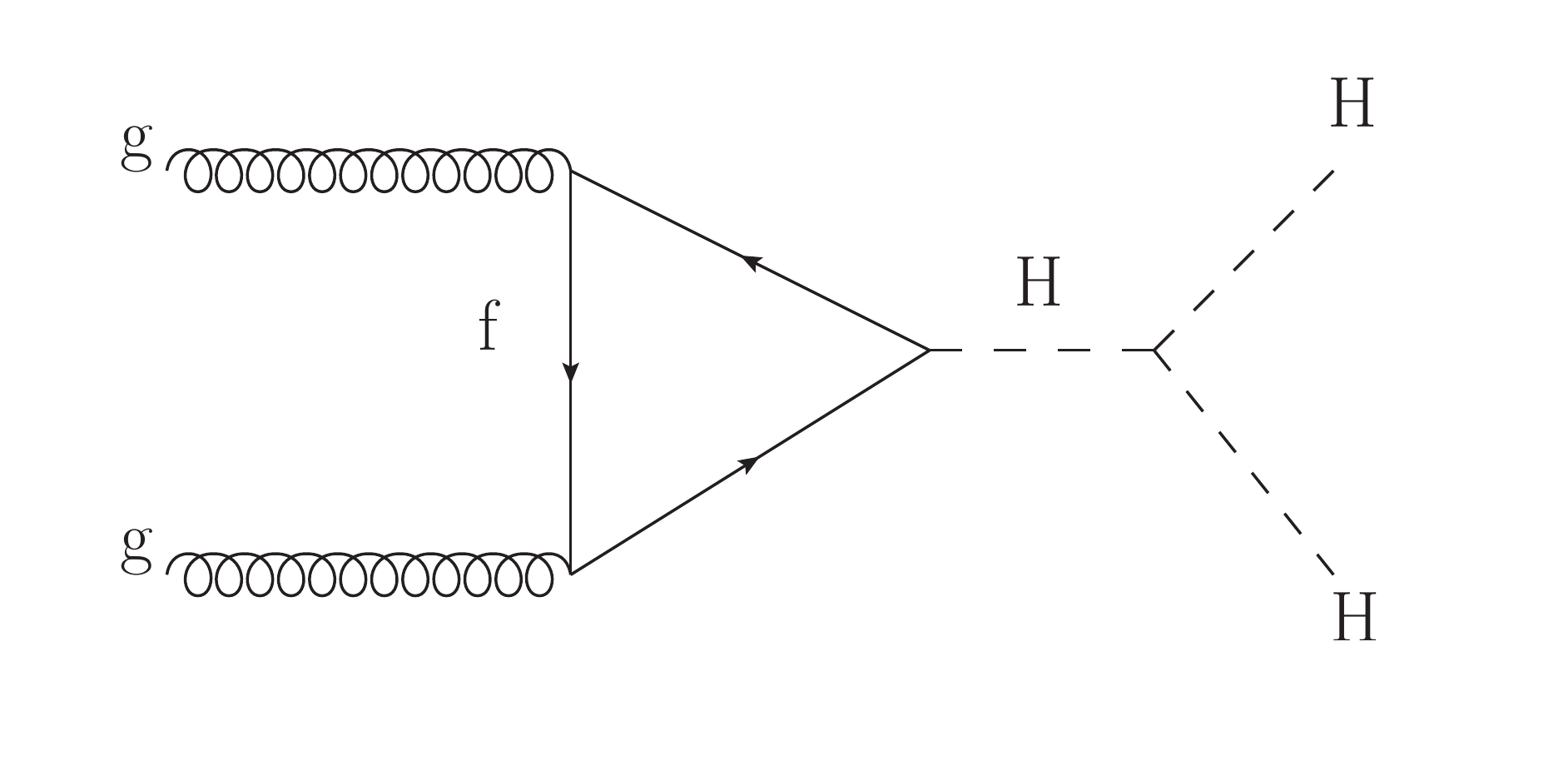}
    \includegraphics[width=0.5\linewidth]{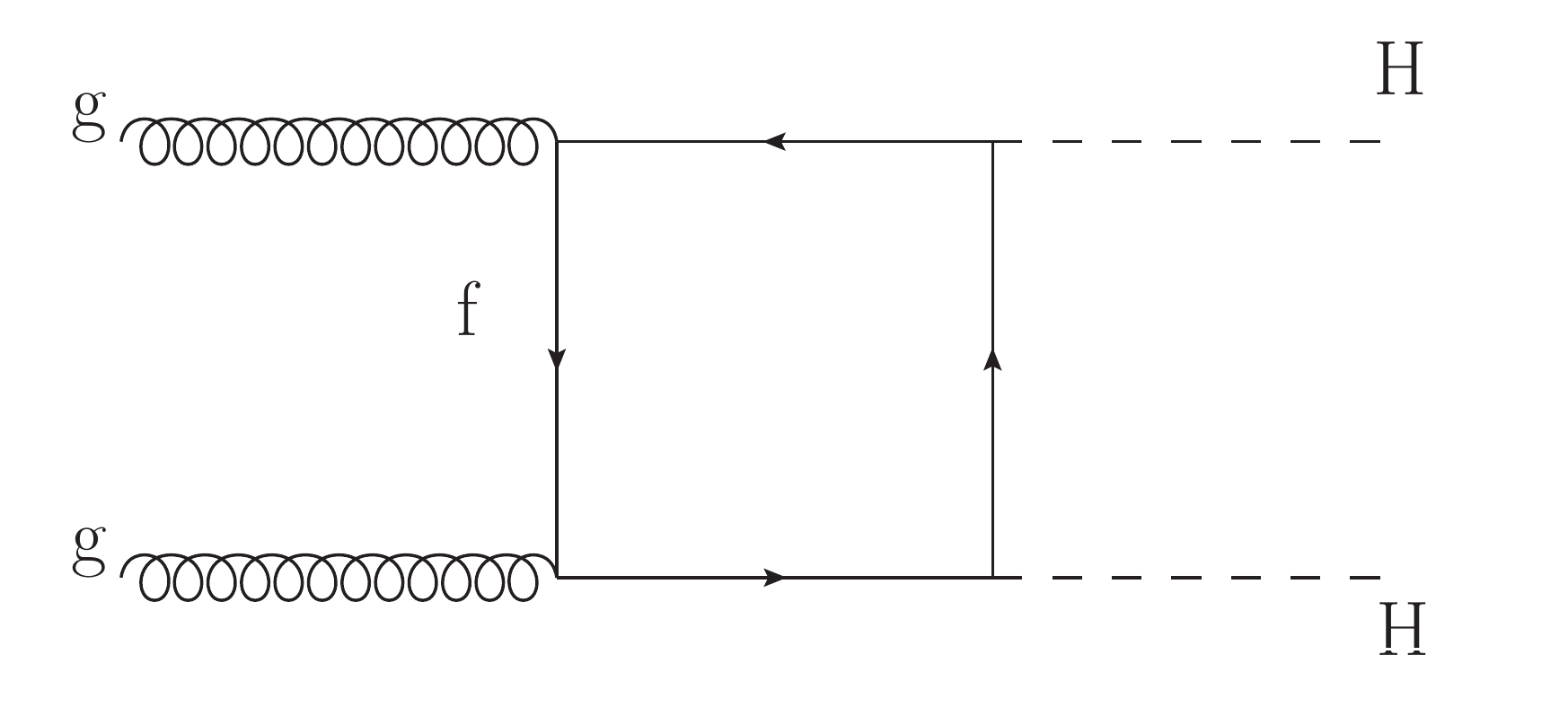}
  \caption{The Higgs pair production diagrams contributing to the
    gluon fusion process at LO are shown, for a generic fermion $f$.}
  \label{fig:HHdiagrams}
\end{figure}

The Higgs boson pair production cross section is dominated by
gluon fusion, as is the single production cross
section~\cite{Plehn:1996wb, Dawson:1998py}. For the
pair production, other modes, like $qq \to qqHH$,$VHH$, $t \bar{t} HH$ are a
factor of 10-30 smaller~\cite{Djouadi:1999rca,Gianotti:2002xx,
  Moretti:2004wa, Baglio:2012np}, and
thus we do not consider them in the rest of our analysis.
At leading order (LO), there are two main contributions: a diagram containing
a `triangle' loop, and one containing a `box' loop of heavy quarks, as
shown in Fig.~\ref{fig:HHdiagrams}. By far the most dominant
contribution comes from the top quark loops, with a smaller
sub-dominant bottom quark contribution. The production of a
single, on-shell Higgs boson only contains a diagram of the `triangle'
type. The triangle diagram can
only contain initial-state gluons in a spin-0 state, whereas the box
contribution can contain both spin-0 and spin-2
configurations. Therefore, there are two Lorentz structures involved
in the box diagram matrix element. At LO, we may write, schematically:
\begin{equation}\label{eq:sigmaHHschem}
\sigma_{HH}^{LO} = \lvert \sum_q ( \alpha_q C^{(1)}_{q,\mathrm{tri}} + \beta_q C^{(1)}_{q,\mathrm{box}})
\rvert^2 +\lvert \sum_q  \gamma_q  C^{(2)}_{q,\mathrm{box}} \rvert^2 \;,
\end{equation}
where $C_{q,\mathrm{tri}}^{(1)}$ represents the matrix element for the
triangle contributions and $C_{q,\mathrm{box}}^{(i)}$ represents the matrix
element for the two Lorentz structures ($i = 1,2$) coming from the box
contributions~\cite{Glover:1987nx, Plehn:1996wb}, for each of the quark flavours $q = \{t,b\}$.

The parameters $\alpha_q$, $\beta_q$ and $\gamma_q$ for quark flavour $q$ are given in terms of the Standard Model Lagrangian parameters by:
\begin{eqnarray}\label{eq:sigmadefs}
\alpha_q = \lambda y_q\;, \nonumber \\
\beta_q = \gamma_q = y_q^2 \;,
\end{eqnarray}
where $q =\{ t,b\}$, $\lambda$ is the (normalised) Higgs triple coupling defined in the previous
section and $y_q$ is the normalised $Hq\bar{q}$ coupling (after electroweak symmetry breaking and assumed to be real) defined with
respect to the SM value: $y_q \equiv Y_q/Y_q^{\mathrm{SM}}$ ($Y_q$
being the resulting coupling and $Y_q^{\mathrm{SM}}$ the SM value). In contrast, the
single Higgs cross section, again, schematically, will only contain
the matrix element squared $\lvert\sum_q C^{(1)}_{q,\mathrm{tri}}\rvert^2$.

We have performed numerical fits using the results of the \texttt{hpair} program~\cite{hpair}, used to
calculate the total cross section for Higgs boson pair
production at leading and approximate next-to-leading (NLO)
orders. The fits were done employing MSTW2008lo68cl and
MSTW2008nlo68cl parton density functions~\cite{Martin:2009iq} and
using top and bottom quark masses of 174.0~GeV and 4.5~GeV
respectively. We have obtained:
\begin{eqnarray}\label{eq:sigmaHHfit}
\sigma^{\rm LO}_{HH} [\mathrm{fb}] = 5.22 \lambda^2 y_t^2 - 25.1 \lambda y_t^3 + 37.3 y_t^4 + {\cal O} (\lambda Y_b y_t^2) \;,\nonumber \\
\sigma^{\rm NLO}_{HH} [\mathrm{fb}] = 9.66 \lambda^2 y_t^2 - 46.9 \lambda y_t^3 + 70.1 y_t^4 + {\cal O} (\lambda Y_b y_t^2) \;,
\end{eqnarray}
where we are not showing terms suppressed by the (un-normalised) $Hb\bar{b}$ coupling, $Y_b$. In fact, we have checked
explicitly that a fit performed ignoring the bottom quark terms
results in form factors accurate at the 1\% level and a total cross
section accurate to better than the 0.2\% level (within the SM). Thus,
for simplicity, we neglect the bottom contributions in the discussion
that follows in the rest of this section. We do, however, include the
bottom quark loops in our numerical results throughout this paper. 

It is evident from Eqs.~(\ref{eq:sigmaHHschem})-(\ref{eq:sigmaHHfit})  that the Higgs pair
production cross section contains an interference term proportional to
$(\lambda y_t^3)$. Hence, for positive values of $(\lambda y_t^3)$ the
cross section is reduced, whereas for negative values, it is
enhanced. The box squared term is dominant, and scales as $y_t^4$,
whereas the triangle squared term is subdominant due to the off-shell
Higgs boson which then decays to Higgs boson pairs, and scales as
$\lambda^2 y_t^2$. Also note that there exists a minimum
value of $\sigma^{\rm NLO}_{HH}$ at $\lambda = \lambda_\mathrm{min}
\simeq 2.46 y_t$ (taking into account the bottom quark
contributions). The cross section $\sigma_{HH}$ is symmetric about the
point $\lambda_\mathrm{min}$.

We note that the above structure, and hence the different contributions to
the cross section, can of course be modified if new physics that allows new resonances to run
in the triangle and box loops (or adds new couplings, like an $ff HH$ interaction) is
present~\cite{Grober:2010yv,Contino:2012xk,Kribs:2012kz,Gillioz:2012se,Dawson:2012mk,Dolan:2012ac}. For
simplicity, in the present article we will focus on the Standard Model
itself, as well as scenarios where the possible higher-dimensional
operators, induced by such new physics, are subdominant with respect to changes in the $\lambda$ and $y_t$ couplings. 

Examples of such scenarios would be models where a Higgs boson $H$ mixes with
another scalar $S$, like in Higgs Portal~\cite{Patt:2006fw, Schabinger:2005ei} or Two-Higgs Doublets Models~(see, e.g.~\cite{Branco:2011iw}),
where no new particles run in the loop. Here the pair production cross
section of the SM-like Higgs boson $H$ will get modified only by having a resonant effect in the s-channel
diagram, due to the new scalar.\footnote{Even if new coloured fermions are present, their
  contribution can be neglected if their couplings are small or if
  they are very heavy and decouple.} Indeed, one can obtain a 10-20 \%
change in $y_t$ and arbitrary values for $\lambda$, together with a
negligible resonant contribution, by selecting
appropriately the free parameters that appear in such
theories.\footnote{In specific examples we have found that one can
  arrange to have a heavy $S$ particle with a small $SHH$ coupling,
  such that its resonance effect will not affect the SM-like Higgs
  pair production rate, and with a moderate deviation in the
  respective $HHH$ coupling. The price to pay
  for $S$ being heavy is to have the other trilinear scalar couplings, $SSH$ and $SSS$ to
  be ${\cal O}(1)$, but still consistent with the perturbativity
  condition, $\lambda \ll \sqrt{4 \pi}$.} The new scalar $S$ may be outside of LHC reach if it is sufficiently heavy, or with reduced couplings to SM
particles (see, e.g.~\cite{Gupta:2012mi}). Even if the new scalar particle is observed,  the measurement of the parameter $\lambda$ will still be a meaningful and interesting question.

\section{Ratios of cross sections}\label{sec:ratios}
It has been pointed out in Ref.~\cite{Djouadi:2012rh} that the ratio of cross sections
between Higgs pair production and single Higgs production:
\begin{equation}
C_{HH} = \frac{ \sigma(gg \rightarrow HH) } { \sigma(gg \rightarrow H)
} \equiv \frac{ \sigma_{HH} }{ \sigma_H },
\end{equation}
could be more accurately determined theoretically than the Higgs-pair production
cross section itself.\footnote{Note that a somewhat different, but
  related, idea of taking ratios of cross sections for various processes at different energies was explored
  in~\cite{Mangano:2012mh}.} This is based on the fact that the processes are both
gluon-initiated and the respective higher-order QCD corrections could be very
similar. Hence, it is assumed that a large component of the QCD uncertainties drop out in the
ratio $C_{HH}$. Moreover, experimental systematic uncertainties that affect both
cross sections may cancel out by taking the ratio. An example is the
luminosity uncertainty, which should cancel out provided the same
amount of data is used in both measurements. 

\begin{figure}[!htb]
  \centering
    \includegraphics[width=0.9\linewidth]{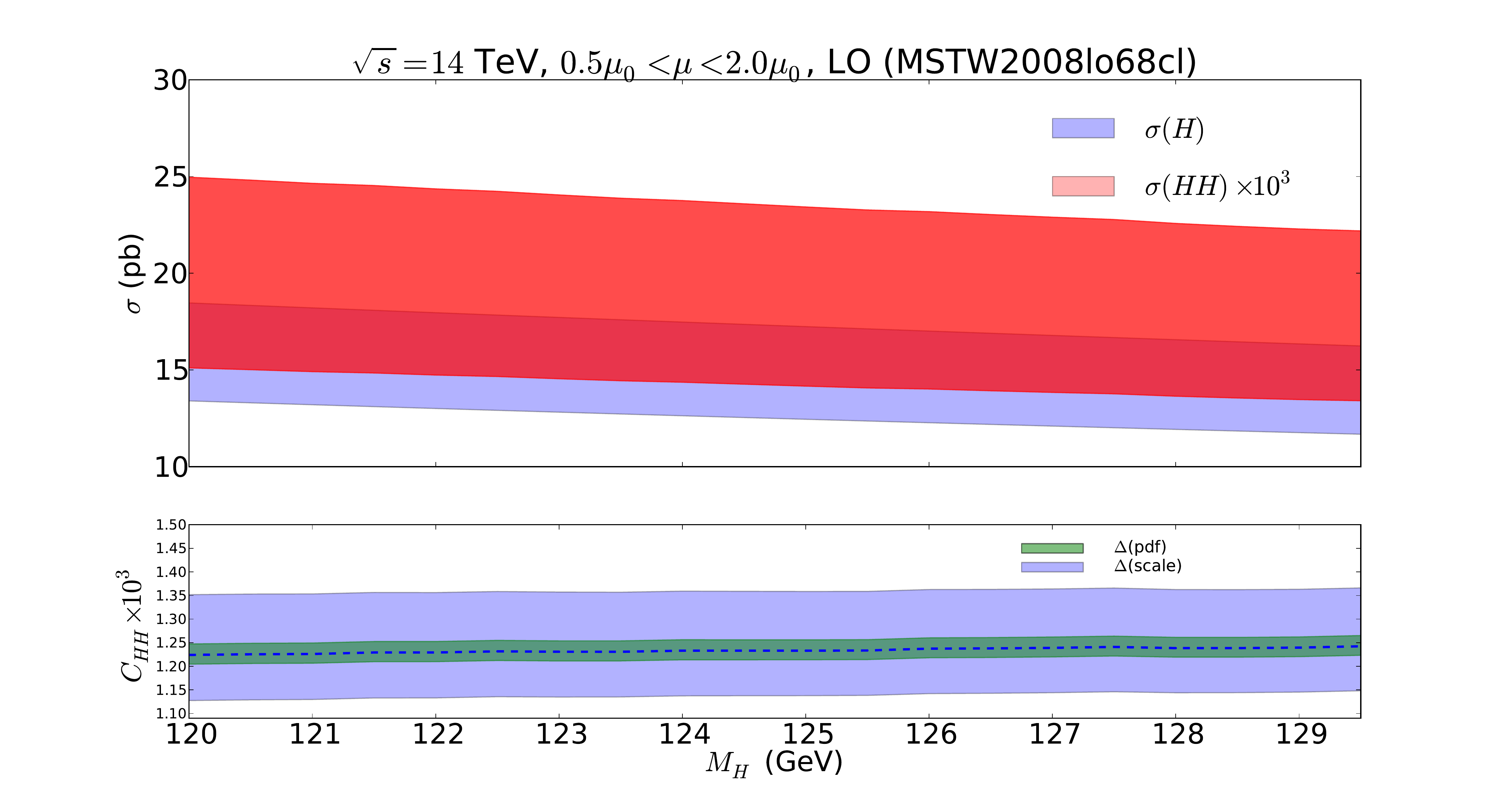}
  \caption{The cross sections for single and double Higgs boson production
    at leading order using the MSTW2008lo68cl PDF set. In the lower
    plot, the fractional uncertainty due to scale variation is shown
    in the blue band, as well as the PDF uncertainty in the green band.}
  \label{fig:xsectionslo}
\end{figure}

\begin{figure}[!htb]
  \centering
    \includegraphics[width=0.9\linewidth]{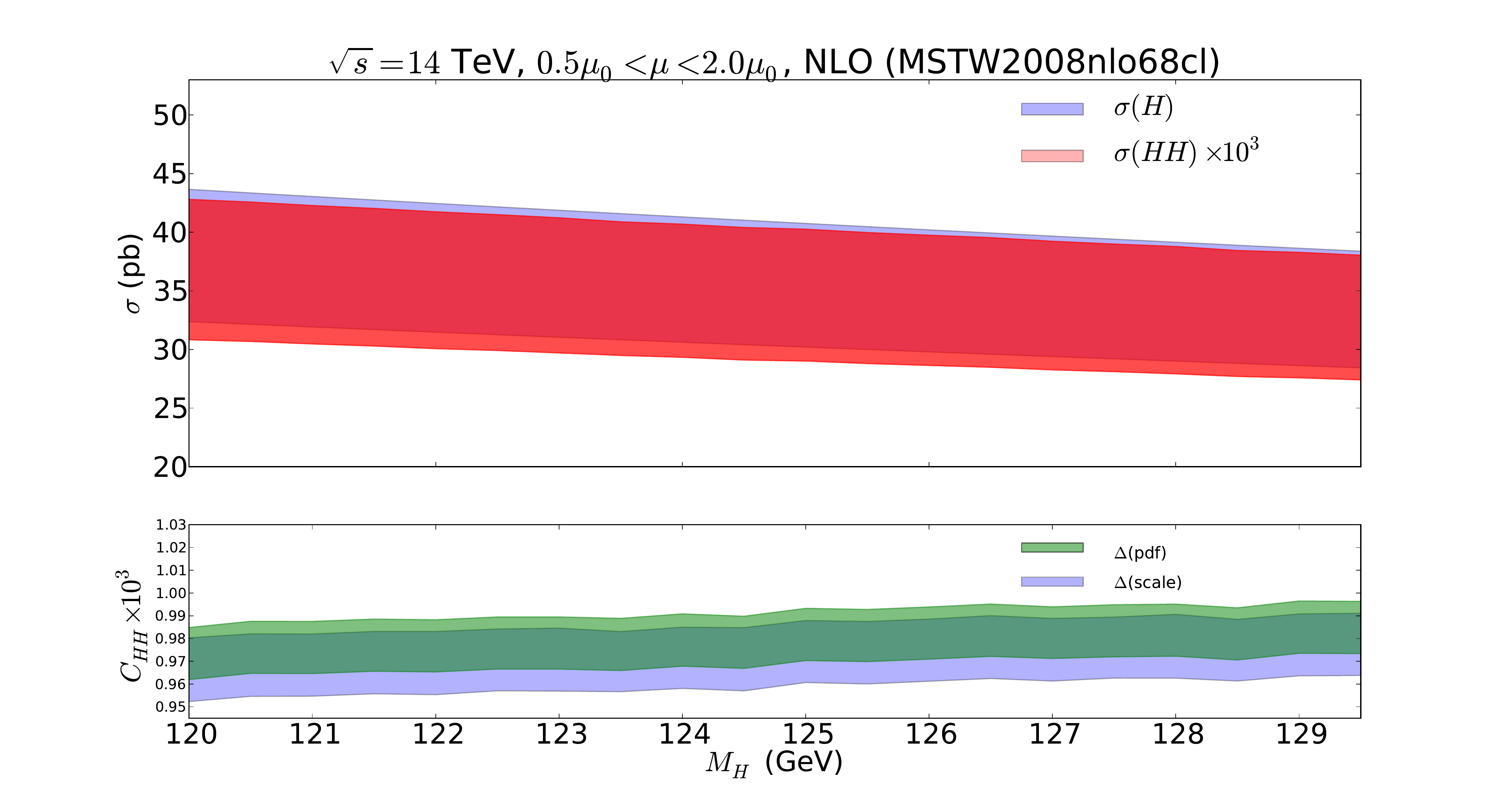}
  \caption{The cross sections for single and double Higgs boson production at next-to-leading order using the MSTW2008nlo68cl PDF set. In the lower
    plot, the fractional uncertainty due to scale variation is shown
    in the blue band, as well as the PDF uncertainty in the green band.}
  \label{fig:xsectionsnlo}
\end{figure}

Here we investigate the extent to which the above assumptions are
correct, using the available calculations for the cross sections. We
begin by considering the LO and NLO calculations for $\sigma(gg \rightarrow HH)$ and $\sigma(gg
\rightarrow H)$ at the LHC at 14~TeV.\footnote{All calculations in the
  present section have been performed in the SM,
  i.e. $\lambda= 1$ and $y_t = 1$. We do not expect the theoretical
  uncertainties to vary substantially with these values, since the variation arises from terms with logarithmic ratios of scales, whose coefficients are often determined by universal QCD functions, namely the $\beta$ function or the Altarelli-Parisi kernels, depending on whether the renormalization or factorization scale is involved. } Using the MSTW2008lo68cl and
MSTW2008nlo68cl parton density functions~\cite{Martin:2009iq}, we
show in Figs.~\ref{fig:xsectionslo} and~\ref{fig:xsectionsnlo} the cross sections as well as
their ratios, $C_{HH}$, as a function of the Higgs mass at both LO and NLO.\footnote{It is important to
  note that the NLO calculation for $HH$ production has been performed in the heavy top
  mass limit, and hence it is expected to be approximate. At LO, the accuracy of the large top
  mass approximation is $\mathcal{O}(10\%)$~\cite{Baur:2002rb,
    Binoth:2006ym, Dawson:2012mk}. Note that the sub-dominant effects
  of the bottom quark are kept in the
  calculations throughout the paper where they are available: up to LO
  in $HH$ production and to NLO in single Higgs production.} We present the scale
uncertainty obtained by varying the factorisation and renormalization
scales (set to be equal) between $[0.5~ \mu_0, 2.0~ \mu_0]$, where $\mu_0 = M_H$ for the
\texttt{higlu} program, used to obtain the single Higgs cross
sections~\cite{higlu}, and $\mu_0 = M_{HH}$ for the \texttt{hpair} program (where $M_{HH}$ is the invariant mass of the Higgs pair), used for the Higgs pair production cross
sections~\cite{hpair}. The scale choices are the natural ones for each
of the processes but we verified that the conclusions are not altered
substantially by changing the \texttt{hpair} scale, i.e. the
numerator, to equal the scale that appears in the
denominator, $\mu_0 = M_{H}$. Implicit in the calculation of the scale
uncertainty of the ratio $C_{HH}$, is the fact that the scale variation
of the single and double Higgs cross sections between $0.5 \mu_0$ and
$2.0 \mu_0$ is fully correlated: i.e., we obtain the upper and lower
variations of the ratio by dividing the cross sections with the same
magnitude of variation of the scale. This is an approximation that is justified
since the two processes possess similar topologies, and is in fact one
of the main insights in favour of using $C_{HH}$. We also show, in the ratio, the resulting PDF uncertainty, calculated using the
MSTW2008nlo68cl error sets according to the prescription found in~\cite{Watt:2011kp}.

Several observations on the behaviour of the $C_{HH}$ ratio can be made. First
of all, it is evident that the fractional uncertainty due to scale variation
is reduced with respect to the individual calculations in both leading
and next-to-leading orders: for the LO case, the individual cross
sections have a $\sim \pm 20\%$ (single Higgs boson production) and
$\sim \pm 25\%$
(double Higgs boson production) scale uncertainty, whereas the ratio has a
$\sim \pm 9\%$ scale uncertainty. For the NLO case, it is reduced from
$\sim \pm 17\%$ (single and double Higgs boson production) to $\sim
\pm 1.5\%$ for
the ratio.\footnote{Note that in Ref.~\cite{Shao:2013bz}, threshold
  resummation effects in SM Higgs pair production in soft-collinear effective theory were
  considered. The authors claim a reduction of the scale uncertainty
  to $3\%$. For other resummation studies in single Higgs production see,
  for example~\cite{deFlorian:2009hc, Ahrens:2008nc, deFlorian:2012mx}.}

Furthermore, we can explicitly see that the uncertainty due to the QCD corrections partially cancels out: 
even though the individual K-factors in the cross sections $\sigma_H$
and $\sigma_{HH}$ are large, they are also very similar, both being
$\sim 2$. As a consequence, the central value of the ratio only
decreases by a small amount from $\sim$1.25 to $\sim$1.0 when going from LO to NLO. This is an
indication that higher order corrections are quite likely to change
the ratio by an even smaller fraction than the change from LO to NLO,
when it is considered at NNLO, whereas the single Higgs production cross section has a
K-factor of about $\sim$1.5 when compared to
the NLO calculation~\cite{Dittmaier:2011ti}.\footnote{An equivalent calculation at NNLO does
  not presently exist for Higgs pair production.} These findings
support the idea of employing the fully correlated scale variation
described before as a realistic estimate for the theoretical
error.\footnote{Note that studies of theoretical uncertainties in single and double Higgs production
  can be found, respectively, in Refs.~\cite{Ball:2012wy, Baglio:2012np}.}

The PDF uncertainties for the cross sections themselves are not shown
since they are of the order of a few \% and hence
subdominant. The PDF uncertainty is also sub-dominant in the case of
the LO ratio, as shown in Fig.~\ref{fig:xsectionslo}. In the case of the NLO ratio, the PDF uncertainty becomes
comparable to the scale uncertainty as can be seen in
Fig.~\ref{fig:xsectionsnlo}. Combining the two errors in quadrature
would induce an error of $\pm\mathcal{O}(3\%)$, still smaller than the
$\sim \pm 17\%$ error on the NLO Higgs pair production cross
section. To remain conservative, we will assume that the theoretical
errors on $C_{HH}$ and $\sigma_{HH}$ are $\pm 5 \%$ and $\pm 20 \%$,
respectively, in what follows. 

\section{Constraining the self-coupling}\label{sec:constaining}
In the studies conducted in Refs.~\cite{Baur:2002qd, Baur:2003gp}, the Higgs self-coupling was
constrained using the final states $b\bar{b} \gamma \gamma$, $b \bar{b} \mu^+
\mu^-$ and $W^+ W^- W^+ W^-$ (in the high Higgs mass region). The
constraints were obtained by fitting the visible mass distributions in each
process for the signal and backgrounds. 

Here we choose to follow a different strategy: taking into account the
facts that the different signal channels possess a relatively low
number of events and that the shapes of distributions for the
backgrounds (and even the signal) are not always very well known, we employ only information originating from the rates. Furthermore, we use the
theoretically more stable ratio $C_{HH}$ between the double and single
Higgs production cross sections, examined in the previous section. We focus on luminosities of
600~fb$^{-1}$ and 3000~fb$^{-1}$ that can be respectively obtained by ATLAS and CMS together in the first
long-term 14~TeV run, or by the individual experiments in an even
longer-term run at the same energy. We do not attempt to combine between the individual channels,
as this will require a more detailed study from the experimental collaborations. 

\subsection{Variation with self-coupling and top quark Yukawa}
To quantify the possible region that can be constrained using the
ratio $C_{HH}$, we first examine the behaviour of the cross section
for Higgs pair production and the ratio $C_{HH}$ at 14~TeV, when varying the self-coupling $\lambda$, as well as the top
Yukawa, $y_t$. It is important to consider the variation of the top quark
Yukawa determination, since the production rates of both double and single Higgs
production can be substantially affected. Moreover, the expected accuracy on the top
quark Yukawa is expected to be $\pm \mathcal{O}(15\%)$ at
300~fb$^{-1}$ of LHC data at 14~TeV~\cite{Peskin:2012we}.

\begin{figure}[!htb]
    \includegraphics[width=0.5\linewidth]{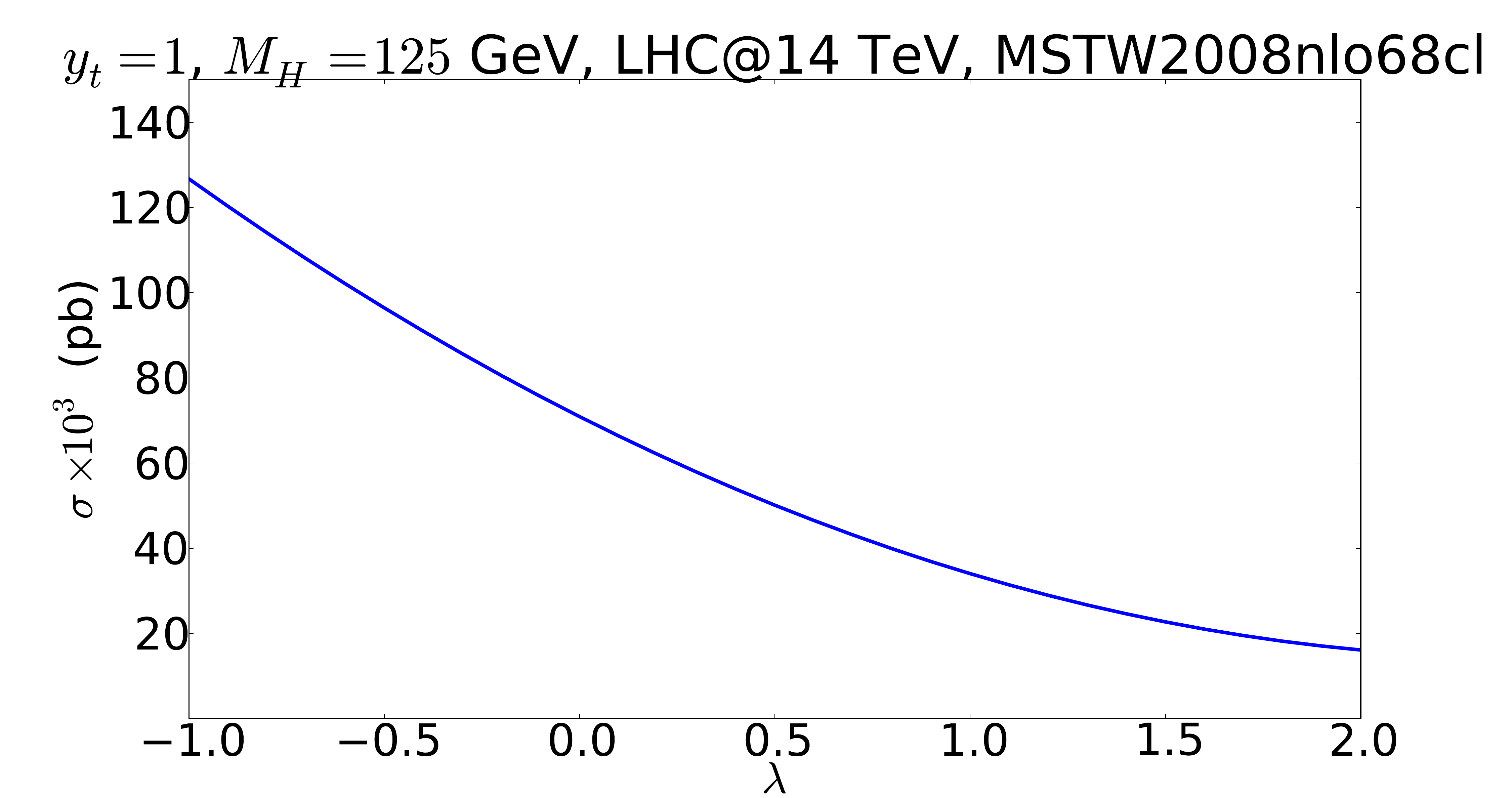}
    \includegraphics[width=0.5\linewidth]{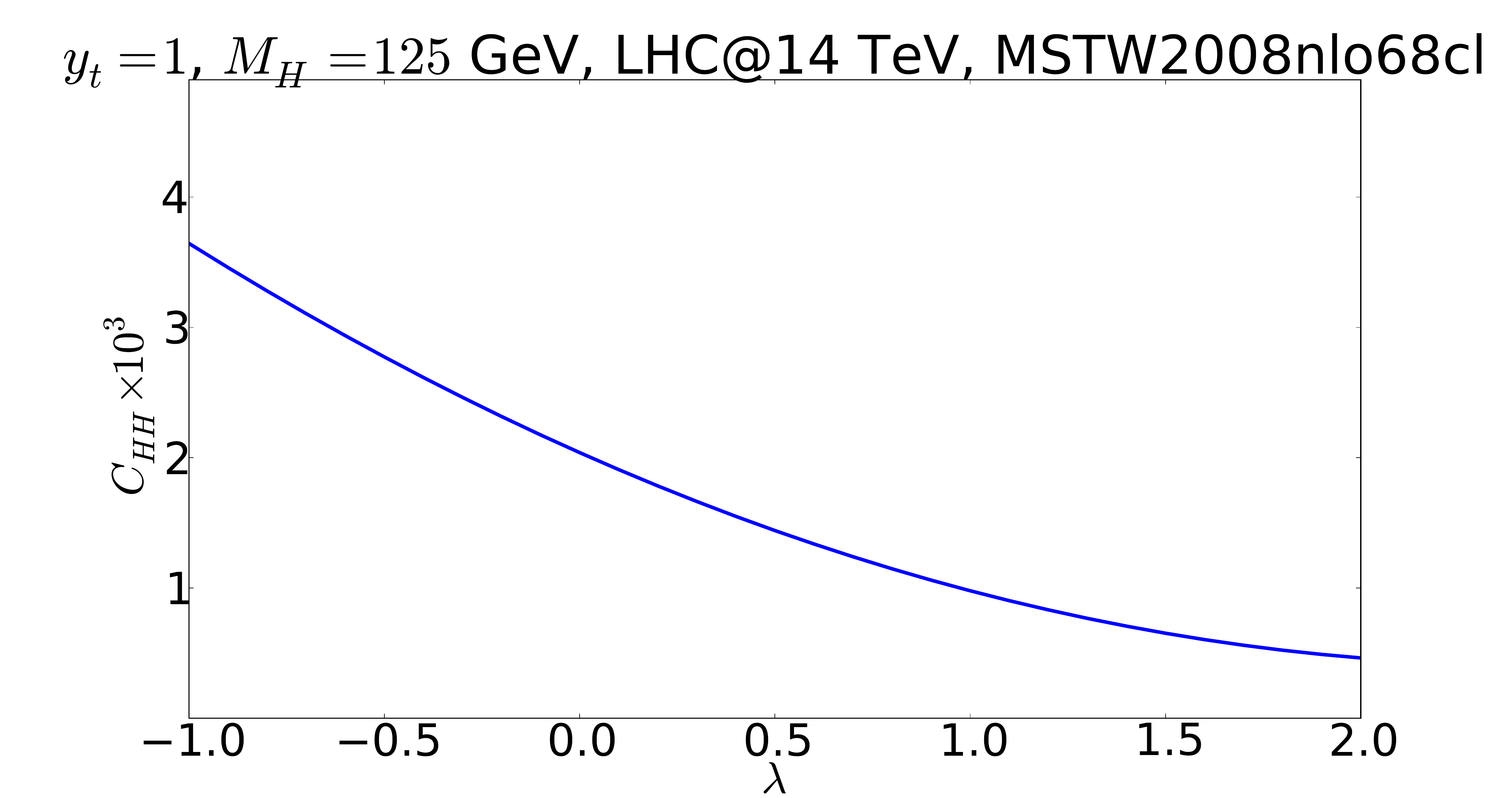}
  \caption{The cross section for double Higgs production and the
    ratio $C_{HH}$ at next-to-leading order using the MSTW2008nlo68cl PDF set, as a
    function of $\lambda$ at $y_t = 1$.}
  \label{fig:flambda}
\end{figure}

\begin{figure}[!htb]
  \includegraphics[width=0.5\linewidth]{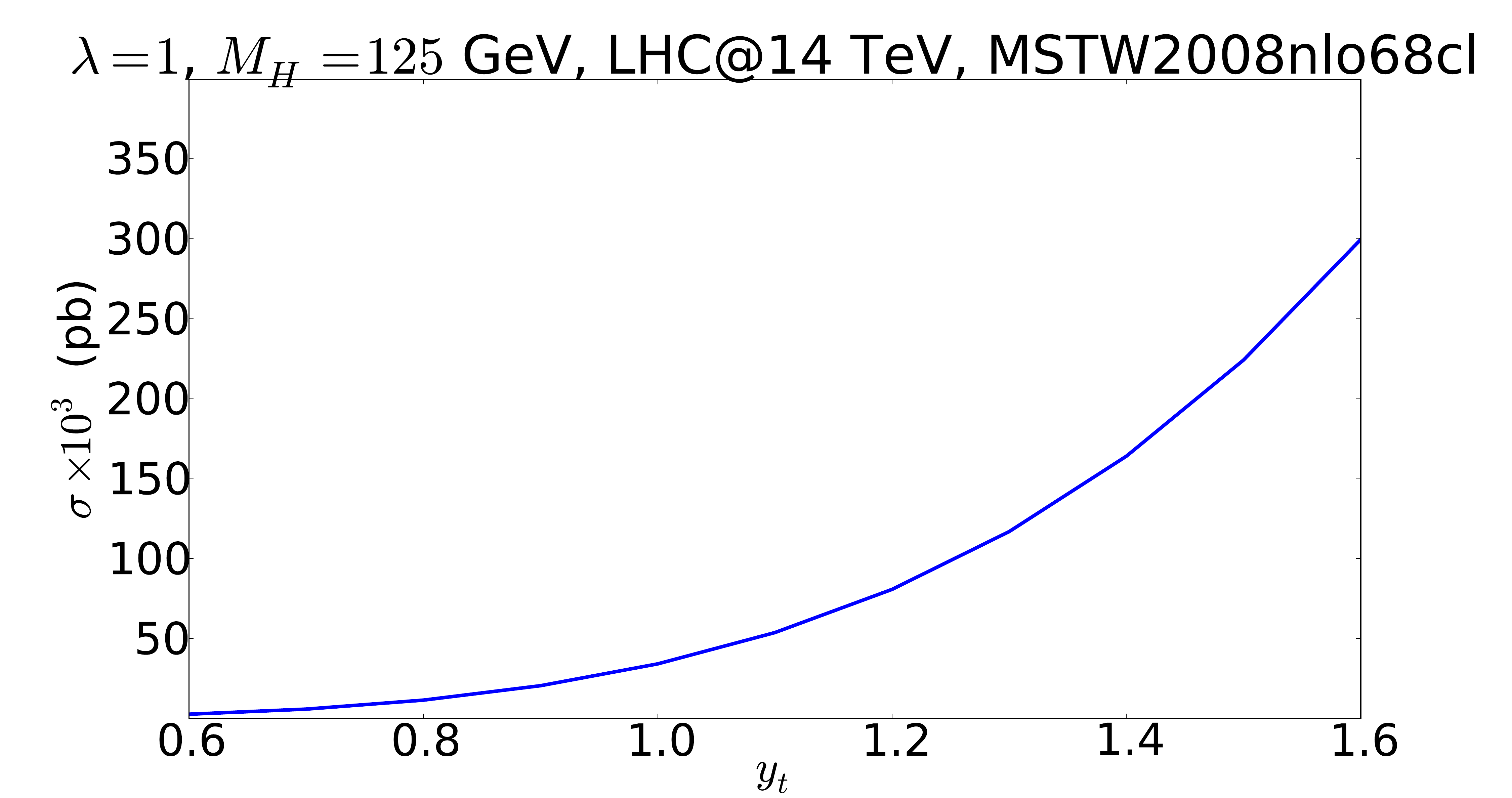}
    \includegraphics[width=0.5\linewidth]{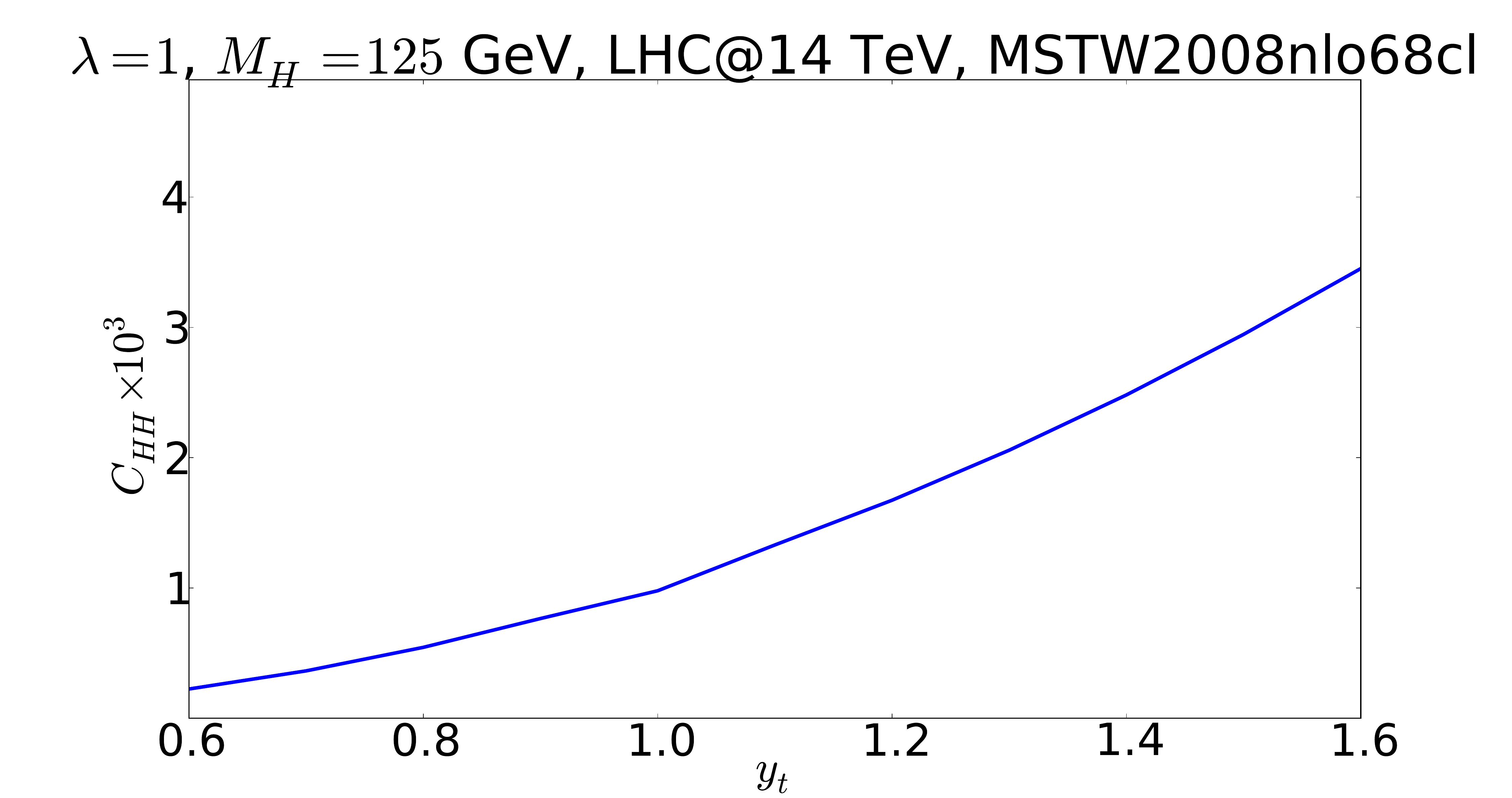}
  \caption{The cross section for double Higgs production and the
    ratio $C_{HH}$ at next-to-leading order using the MSTW2008nlo68cl PDF set, as a
    function of $y_t$ at $\lambda = 1$.}
  \label{fig:fyt}
\end{figure}

We show the cross section $\sigma_{HH}$ and ratio $C_{HH}$ at $y_t = 1$ as a function of $\lambda$, as well as both quantities at $\lambda=1$ as a function of $y_t$ in Figs.~\ref{fig:flambda} and~\ref{fig:fyt}, respectively. Evidently, the effects
of both $\lambda$ and $y_t$ are significant: the cross section varies
from $\sim 30~\mathrm{fb}$ at $(\lambda, y_t) =
(1,1)$ (i.e. the SM values) to $\sim 125~\mathrm{fb}$ at $(\lambda,
y_t) = (-1,1)$ and $\sim 300~\mathrm{fb}$ at $(\lambda, y_t) =
(1,1.6)$. The ratio itself varies from $\sim 10^{-3}$ at $(\lambda, y_t) =
(1,1)$ to $\sim 3.5\times10^{-3}$ at $(\lambda, y_t) =
(-1,1)$ and $(\lambda, y_t) = (1,1.6)$. It is
obvious that negative values of $\lambda$ can be excluded sooner
than the positive values, since the cross section and ratio of cross
sections both increase fast with decreasing $\lambda$.

We note that negative values of $y_t$ are currently viable~\cite{Espinosa:2012im} and physical, and could arise in beyond-the-SM physics models. Since Higgs
pair production only depends on the sign of the product $(\lambda y_t)$,
the corresponding values for $y_t < 0$, $\lambda > 0$ are equivalent
to those for the points with the same absolute values of the
parameters but $y_t > 0$, $\lambda < 0$.\footnote{Note that the
  degeneracy with respect to the sign of $y_t$ that appears in Higgs pair production may be resolved
  through the study of different processes long before the Higgs
  self-coupling is probed. See, for example, Refs.~\cite{Farina:2012xp,Biswas:2012bd}.}

\subsection{Assumptions for experimental uncertainties}
The ratio $C_{HH}$ can be used to derive the expected constraints that
can be obtained at a 14~TeV LHC for different physics models,
including the SM. Certain assumptions on the systematic
uncertainties need to be made for the branching ratios related to each
mode. We first define the following quantities:
\begin{eqnarray}
\sigma_{HH}^{b\bar{b} xx} &\equiv& \sigma_{HH} \times
2\times\mathrm{BR}(b\bar{b}) \times \mathrm{BR}(xx) \;,
\nonumber \\
\sigma_{H}^{b\bar{b}} &\equiv& \sigma_H \times \mathrm{BR}(b\bar{b})\;, 
\end{eqnarray}
where $xx$ denotes the $H\rightarrow xx$ decay mode in question. Hence, we can
invert the above relations to obtain:
\begin{equation}\label{eq:chhexp}
C_{HH}^\mathrm{exp.} = \left. \frac{\sigma_{HH}^{b\bar{b}xx}} { 2 \times
  \sigma^{b\bar{b}}_{H} \times BR(xx) } \right|_{\mathrm{exp.}}\;,
\end{equation}
which is the experimental measurement of the theoretical quantity
$C_{HH}$. 

Since the scope of this article is not a detailed experimental study,
we now make several assumptions on the measurement uncertainties for
each of the quantities in the ratio of Eq.~(\ref{eq:chhexp}). We focus
on the region $\lambda \in (-1.0, \sim2.46)$, since the cross section
is symmetric with respect to the minimum at $\lambda \simeq 2.46$. According to Ref.~\cite{cmseuropean}, the branching ratio of $H \rightarrow
b\bar{b}$ times the cross section for single Higgs is expected to be
known to $\pm 20\%$ after
300~fb$^{-1}$ of data at 14 TeV, and hence we assume that the uncertainty on
$\sigma_H^{b\bar{b}}$ is $\pm 20\%$. Similarly, according
to~\cite{cmseuropean}, the uncertainties on
$\mathrm{BR}(\tau^+\tau^-)$, $\mathrm{BR}(W^+ W^-)$ and
$\mathrm{BR}(\gamma \gamma)$ are expected to be $\pm12\%$, $\pm12\%$ and $\pm16\%$,
respectively, at 300~fb$^{-1}$. To remain conservative, we assume that going beyond
300~fb$^{-1}$ of luminosity, there will be \textit{no} improvement
on these uncertainties. This can be true, for example, if the
measurements are dominated by systematic uncertainties that cannot be
improved further. Moreover, the uncertainty on the cross section of the
measured final state, $\Delta
   \sigma_{HH}^{b\bar{b} xx}$, is estimated by assuming that the
Poisson distribution of the obtained number of events can be
approximated by a Gaussian, for simplicity. Hence, if we expect a
number of $B$ background events and we experimentally measure $N$
events, the error on the signal estimate, $S = N - B$, is given by $\Delta S =
\sqrt{ N + B }$. The expected number of events for the studies we
consider below were taken from~\cite{Baur:2003gp, Dolan:2012rv
  ,Papaefstathiou:2012qe}. We combine all
the estimates of the uncertainties in quadrature for each mode to
obtain an estimate of the total uncertainty:
\begin{equation}
\left(\frac{\Delta C_{HH}}{C_{HH}}\right)^2 = \left(\frac{\Delta
   \sigma_{HH}^{b\bar{b} xx} }{\sigma_{HH}^{b\bar{b} xx}}\right)^2 + \left(\frac{\Delta
    \mathrm{BR} (xx) } { \mathrm{BR} (xx)  }\right)^2 +
\left(\frac{\Delta \sigma_{H}^{b\bar{b}}} {\sigma_{H}^{b\bar{b}} }\right)^2\;. 
\end{equation} 
In what follows we also add the theoretical error estimates in quadrature
to the above. 

\subsection{Deriving constraints}

The ratio of cross sections considered in Section~\ref{sec:ratios}
was calculated under the assumption of validity of the SM. In general,
if one wishes to use the ratio to perform a study of a
different model with a given set of parameters $\{p_i\}$, one
should first:
\begin{itemize}
\item{Calculate the ratio $C_{HH}$ and the corresponding theoretical error as a
    function of the set of parameters $\{p_i\}$. The set $\{p_i\}$
    may, for example, include the new masses and couplings of the theory or
  coefficients of new higher-dimensional operators. }
\item{Estimate, as well as possible, the expected experimental errors
    arising from the measurements of the different components that
    comprise the experimental value of the ratio
    $C_{HH}^\mathrm{exp.}$, as we have done in the previous section.}
\end{itemize} 
With the above at hand, one can then form the following question:
\begin{quote}
Given an assumption for the `true' value of a subset of the model parameters, what is the constraint we \textit{expect} to
impose on these parameters through Higgs pair production?
\end{quote}
Following the above framework, here we perform a study of a simplified
model, which we present as an example of an implementation of the above steps. Thus, we
consider a situation where the Standard Model is valid almost
everywhere, except that we allow the variation of the parameters
$\{p_i\} = \{\lambda, y_t\}$. As we have already discussed at the end of Section~\ref{sec:cross}, such situations may
arise in Higgs Portal or Two-Higgs Doublets Models. Furthermore, in
the same framework, this simplified model will also provide us with limits
on the determination of $\lambda$ within the SM, by setting the `true'
values of $\lambda$ and $y_t$, $\lambda_{\mathrm{true}} = 1 = y_{t,\mathrm{true}}$.

We start by fixing the value of the top Yukawa in this simplified
model to be $y_t = y_{t,\mathrm{true}}=1$. Thus, to answer to the above
question we produce an `exclusion' plot, calculated by drawing the
curves that result in expected measurements that are one or two standard deviations away from the central
value of $C_{HH}$, which is assumed to be equal to that given by $\lambda_\mathrm{true}$. By virtue of
this definition, it is obvious that the central value itself is, of
course, not expected to be excluded. Equivalent plots in this model
can be constructed, by fixing $\lambda_{\mathrm{true}}$ and varying
$y_{t,\mathrm{true}}$, but we do not perform these here. 

\begin{figure}[!htb]
  \centering
    \includegraphics[width=0.8\linewidth]{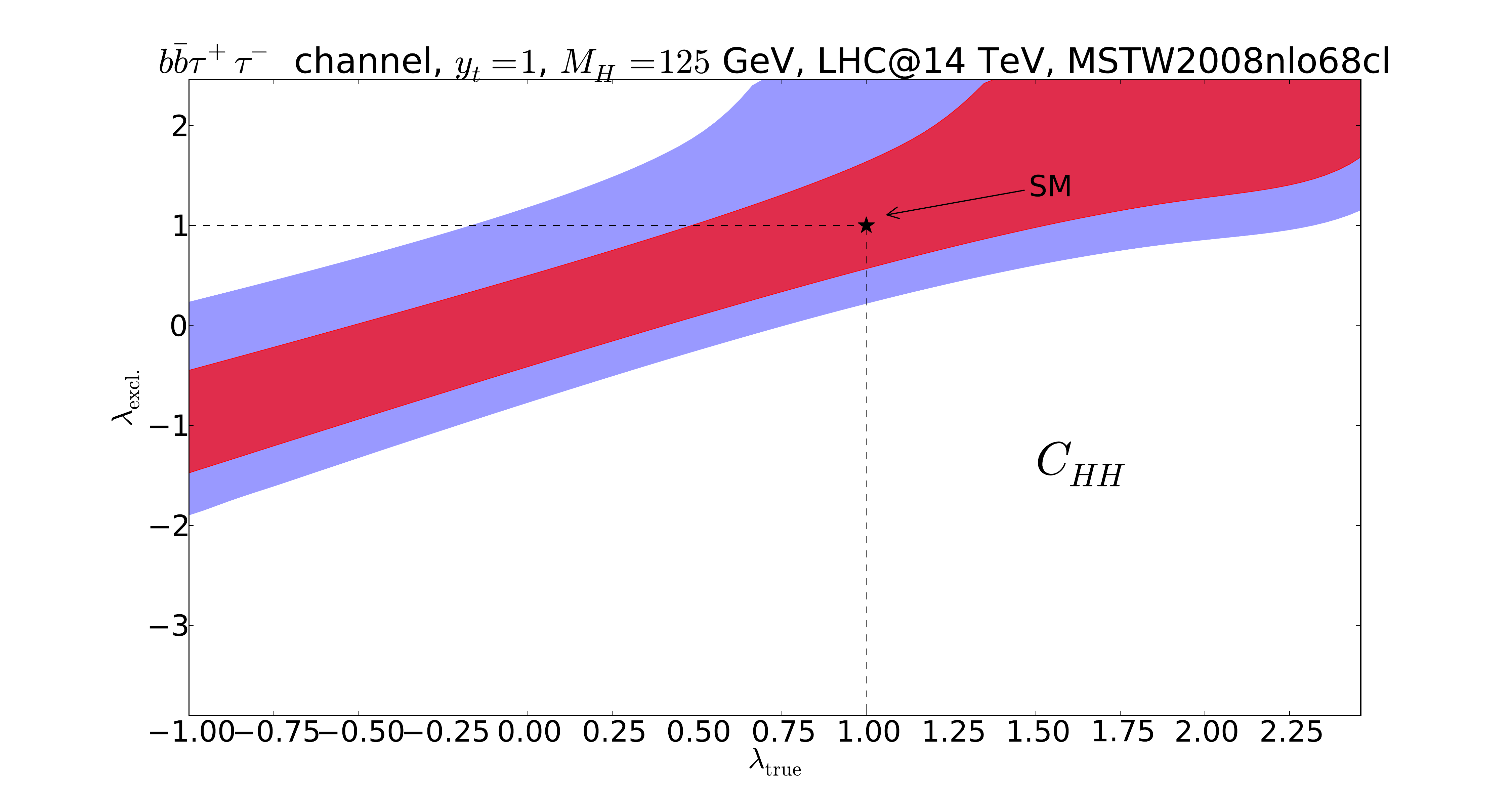}
    \includegraphics[width=0.8\linewidth]{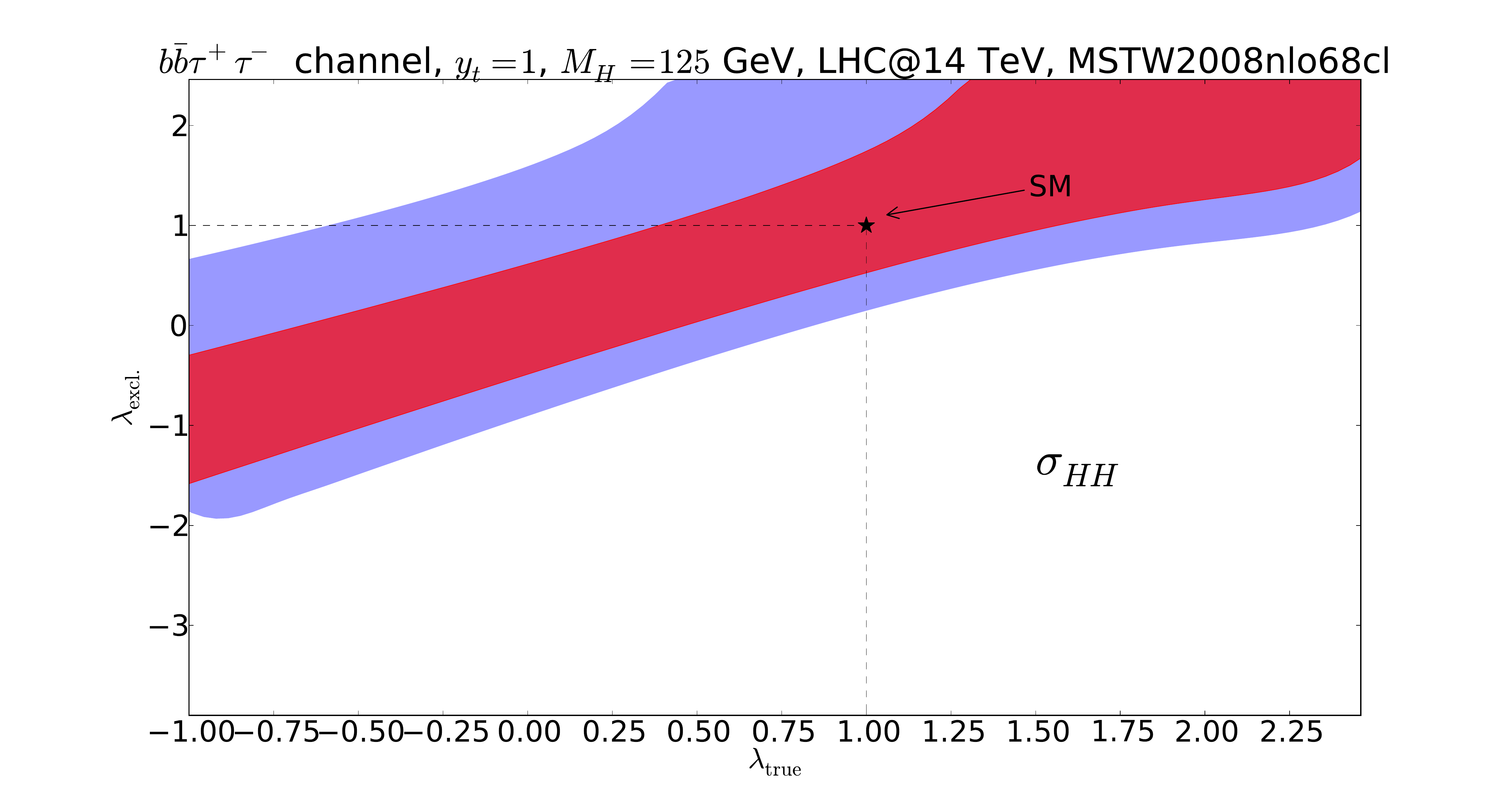}
  \caption{The expected exclusion for $\lambda$ in the simplified model we are considering, at one and two standard deviations
    for a given value of $\lambda_{\mathrm{true}}$ at 600~fb$^{-1}$
    for the $b\bar{b} \tau^+\tau^-$ decay mode. The exclusion
    constructed from the ratio, $C_{HH}$, is shown on the top panel, whereas the
    exclusion obtained from the cross section, $\sigma_{HH}$, is shown
    on the bottom panel. We only show the region up to the symmetric
    minimum at $\lambda\simeq 2.46$. }
  \label{fig:exctautau}
\end{figure}

\begin{figure}[!htb]
\centering
    \includegraphics[width=0.8\linewidth]{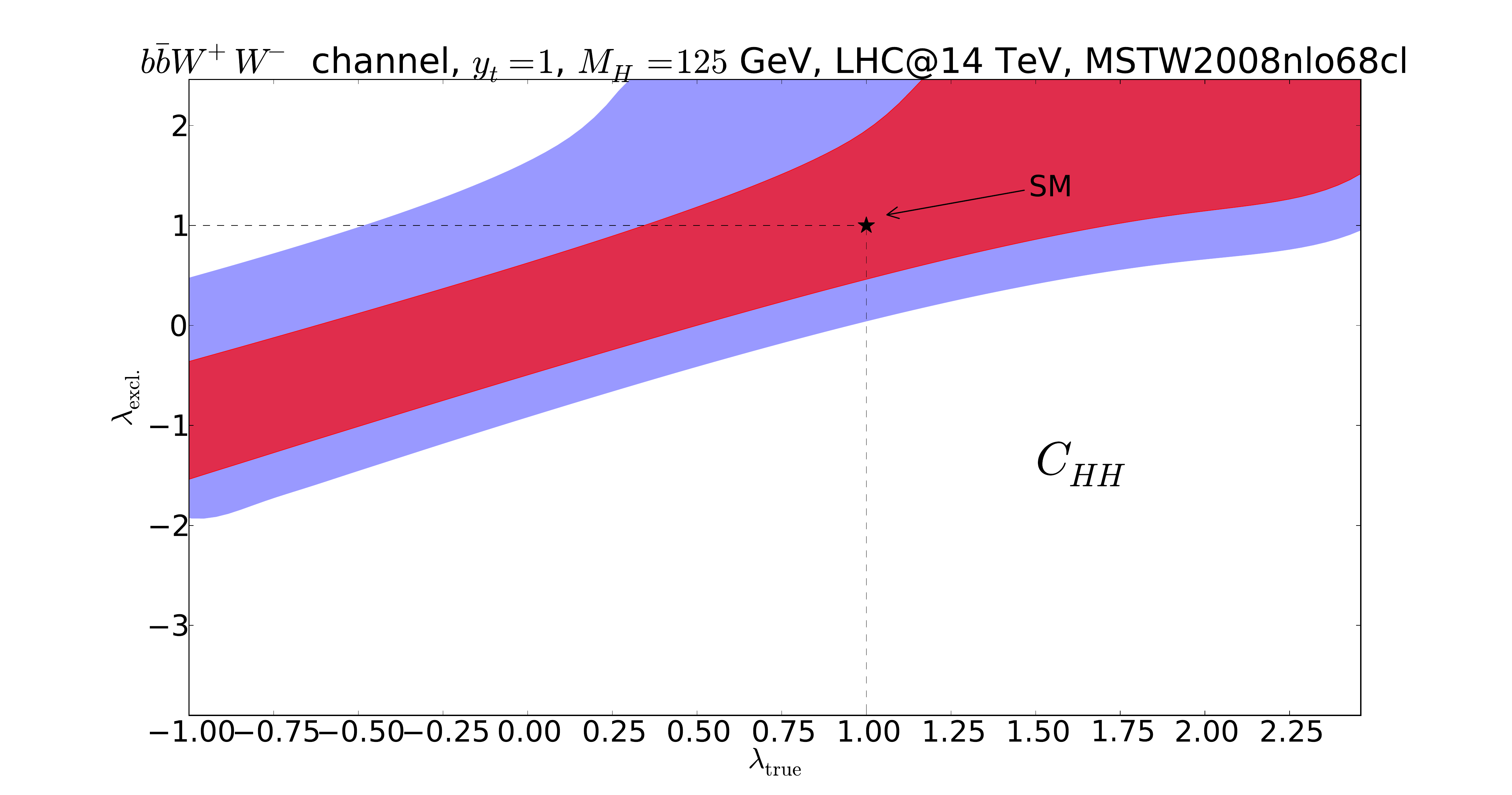}
  \caption{The expected exclusion in the simplified model we are considering, for $\lambda$ at one and two standard deviations
    for a given value of $\lambda_{\mathrm{true}}$ at 600~fb$^{-1}$
    for the $b\bar{b} W^+W^-$ decay mode, constructed by using the
    ratio of cross sections $C_{HH}$. We only show the region up to the symmetric
    minimum at $\lambda \simeq 2.46$. } 
  \label{fig:excww}
\end{figure}

\begin{figure}[!htb]
\centering
    \includegraphics[width=0.8\linewidth]{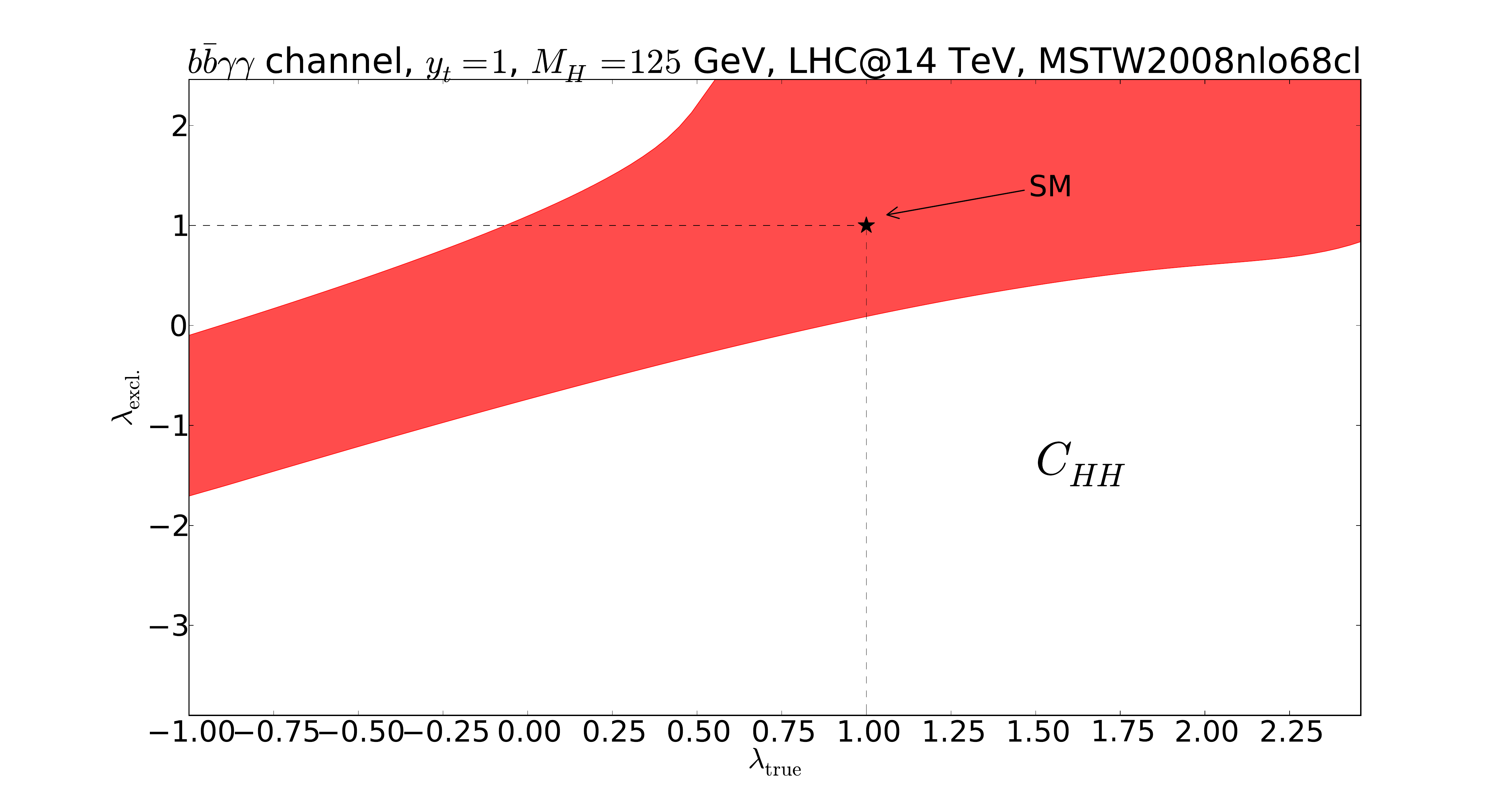}
  \caption{The expected exclusion in the simplified model we are considering, for $\lambda$ at one standard
    deviations for a given value of $\lambda_{\mathrm{true}}$ at 600~fb$^{-1}$
    for the $b\bar{b} \gamma \gamma$ decay mode, constructed by using the
    ratio of cross sections $C_{HH}$. The two standard
    deviations exclusion is not shown since it is weak. We only show the region up to the symmetric
    minimum at $\lambda \simeq 2.46$. }
  \label{fig:excgg}
\end{figure}

Using $C_{HH}$ we draw such curves for 600~fb$^{-1}$ of data in
Figs.~\ref{fig:exctautau},~\ref{fig:excww} and~\ref{fig:excgg} for the
final states $b\bar{b} \tau^+
\tau^-$, $b\bar{b} W^+ W^-$ and $b\bar{b} \gamma \gamma$,
respectively. To bring the three channels to an equal footing, we have rescaled the $b\bar{b}\tau^+
\tau^-$ cross section in~\cite{Dolan:2012rv} by employing a factor of $32.4/28.4$
accounting for the central value of the NLO production cross section
used in~\cite{Papaefstathiou:2012qe}, and moreover, rescaled
by $0.7^2/0.8^2$ for a reduced $\tau$-jet tagging efficiency. For the
$b\bar{b}W^+W^-$ mode in~\cite{Papaefstathiou:2012qe} we also include the tauonic
decays of the $W$ bosons, and for the $b\bar{b} \gamma \gamma$ result in~\cite{Baur:2003gp} we
average between the `hi' and `lo' LHC results to get $6$ versus $12.5$
events at 600~fb$^{-1}$.\footnote{The `hi' and `lo' refer to, respectively, the
  conservative and optimistic assumptions made in~\cite{Baur:2003gp} for the jet to photon misidentification probability.} We have not rescaled the  $b\bar{b} \gamma
\gamma$ analysis, since this was done for a Higgs of mass 120~GeV
in~\cite{Baur:2003gp}. In the lower panel of Fig.~\ref{fig:exctautau}  we also
show the exclusion regions extracted by using the Higgs pair
production cross section measurement itself, with an associated
uncertainty of $\pm 20\%$. We assume that the uncertainty on
$\mathrm{BR}(b\bar{b})$ is the same as that on $\sigma^{b\bar{b}}_H$,
namely $\pm 20\%$. It is obvious that the exclusion obtained from
the cross section is expected to be weaker than that obtained by the
ratio, due to the larger theoretical systematic uncertainty on the cross
section itself. Moreover, the expected exclusion from
$\sigma_{HH}$ will be more affected by experimental systematic
uncertainties which would add to the errors. For completeness, we show the estimated fractional
uncertainty on the ratio, $\Delta C_{HH} / C_{HH}$, used to extract the exclusion regions, for the different processes and investigated luminosities
in Table~\ref{tb:deltachh}. At high luminosity the uncertainties all tend to
  similar numbers since we have assumed that the other contributing
  uncertainties ($\Delta\mathrm{BR} (xx)$ and $\Delta \sigma_{H}^{b\bar{b}}$) do not improve and they become
  systematic-dominated. These values are provided for completeness, as
  an indication, and merit further investigation by the experimental
  collaborations. 
\begin{table}[!htb]
\begin{center}
\begin{tabular}{|l|c|c|c|} \hline
Process & S/B(600~fb$^{-1}$) & $\Delta C_{HH} / C_{HH}$ (600~fb$^{-1}$) &$\Delta C_{HH} /C_{HH}$ (3000~fb$^{-1}$) \\ \hline
$b\bar{b} \tau^+ \tau^-$ & 50/104 & 0.400 & 0.279 \\ \hline
$b\bar{b} W^+ W^-$ & 11.2/7.4 & 0.513 & 0.314  \\ \hline
$b\bar{b} \gamma \gamma$ & 6/12.5 & 0.964 & 0.490 \\ \hline
\end{tabular}
\end{center}
\caption{The table shows expected number of signal (S) and background
  (B) events for SM Higgs pair production, resulting at 600~fb$^{-1}$, and the respective fractional uncertainties on the ratio of double-to-single
  Higgs boson production cross sections, $\Delta C_{HH} / C_{HH}$, for the different channels and
  the two investigated LHC luminosities, 600~fb$^{-1}$ and
  3000~fb$^{-1}$, using $M_H = 125$~GeV. The fractional uncertainties include the
theoretical error due to the scale/parton density functions
uncertainties, assumed to be 5\%.}
\label{tb:deltachh}
\end{table}

The interpretation of the `exclusion' curves is simple: as an example,
if we assume or believe that the `true' value of the triple Higgs
coupling in this model is $\lambda_\mathrm{true} = 1$, then by examining
Fig.~\ref{fig:exctautau} for the $b\bar{b} \tau^+ \tau^-$ mode at
600~fb$^{-1}$, we can conclude that using $C_{HH}$ the expected
experimental result should lie within $\lambda \in (0.57, 1.64)$ with $\sim$68\% confidence level. We expect to exclude any values outside this range
after 600~fb$^{-1}$, given the value $\lambda_\mathrm{true} = 1$.  We show the
collected exclusion limits for $\lambda_\mathrm{true} = 1$ and $y_{t,\mathrm{true}} =
1$ (i.e. the SM values) at $1\sigma$ and $2\sigma$ at 600~fb$^{-1}$
as well as the end-of-run LHC integrated luminosity of 3000~fb$^{-1}$
in Table~\ref{tb:excl}. The 3000~fb$^{-1}$ values have also been calculated
by assuming no improvement in the uncertainty estimates that we have
assumed at 600~fb$^{-1}$. The table demonstrates an important
conclusion: it is possible, using the discovery of the three viable
channels, to constrain the trilinear coupling $\lambda$ in the SM to
be positive at 95\% confidence level at
600~fb$^{-1}$. Moreover,
a naive combination of the `uncertainties', at $1\sigma$ about $\lambda_\mathrm{true}$, over the three
channels indicates that a measurement of accuracy $\sim+30\%$ and $\sim-20\%$ is
possible simply by using the rates at 3000~fb$^{-1}$. Note that the
curves have been drawn up to $\lambda_\mathrm{min} \simeq 2.46$. The
regions beyond that value are determined by the mirror symmetry with
respect to $\lambda_\mathrm{min}$ (the cross section is degenerate for $\lambda \rightarrow 2 
\lambda_\mathrm{min} - \lambda $, which makes those values of $\lambda$
indistinguishable).

We should emphasise at this point that
Figs.~\ref{fig:exctautau},~\ref{fig:excww} and~\ref{fig:excgg} do not
represent the Standard Model, \textit{except} at
$\lambda_{\mathrm{true}} = 1$, and should be taken simply as an
example of the suggested framework in a simplified, but still not
unrealistic, scenario. 

\begin{table}[!htb]
\begin{center}
\begin{tabular}{|l|c|c|c|c|} \hline
Process & 600~fb$^{-1}$ (2$\sigma$) & 600~fb$^{-1}$ (1$\sigma$) &
3000~fb$^{-1}$ 2$\sigma$ &  3000~fb$^{-1}$ 1$\sigma$\\ \hline
$b\bar{b} \tau^+ \tau^-$ &  (0.22, 4.70) & (0.57, 1.64) & (0.42, 2.13) & (0.69, 1.40)\\ \hline
$b\bar{b} W^+ W^-$ & (0.04, 4.88) & (0.46, 1.95)  & (0.36, 4.56)  & (0.65, 1.46)  \\ \hline
$b\bar{b} \gamma \gamma$ &  (-0.56, 5.48) & (0.09, 4.83)& (0.08, 4.84) & (0.48, 1.87)  \\ \hline
\end{tabular}
\end{center}
\caption{The expected limits on $\lambda$ at 1$\sigma$ and 2$\sigma$ confidence
  levels in the Standard Model ($\lambda_\mathrm{true} = 1$, $y_{t,\mathrm{true}} =
  1$). The results have been derived using $C_{HH}$ and are shown for
  600~fb$^{-1}$ and 3000~fb$^{-1}$. Note that there can be either one
  or two regions, in both cases symmetric about the minimum at $\lambda\simeq 2.46$. Where
  there may exist a second valid region, we only show the lower one.}
\label{tb:excl}
\end{table}

It is interesting to compare the regions obtained by the above method
for the SM, with those obtained in
Ref.~\cite{Baur:2003gp}, where the authors used the only viable mode for a low
mass Higgs boson at the time ($M_H = 120$~GeV),
$b\bar{b}\gamma\gamma$, to extract $\lambda$ from the visible mass distribution. After background
subtraction, their best limit at 600~fb$^{-1}$ was $\lambda
\in (0.26, 1.94)$ at 1$\sigma$. Here, for the $b\bar{b}\tau^+\tau^-$
we obtain $\lambda \in (0.57, 1.64)$, for the $b\bar{b}W^+W^-$ mode we obtain $\lambda \in (0.46,1.95)$ and for the $b\bar{b}\gamma\gamma$
mode, $\lambda \in (0.09, 4.83)$, where the latter corresponds to the
full interval, symmetric about the minimum. It is evident
that the ratio provides a comparable exclusion region, especially
considering the fact that Ref.~\cite{Baur:2003gp} considers relatively optimistic background subtraction. However, the
ratio possesses advantages over the distribution analysis that may
contain systematic uncertainties induced by the modelling of the shapes of
both the signal and background. Note that an interesting study of the
theoretical sensitivity of different initial states ($gg \rightarrow HH$, $qq'
\rightarrow HH q q'$, $q\bar{q}'\rightarrow WHH$ and $q\bar{q} \rightarrow
ZHH$) on the trilinear coupling can be found in~\cite{Baglio:2012np}.

\begin{figure}[!htb]
  \centering
    \includegraphics[width=0.8\linewidth]{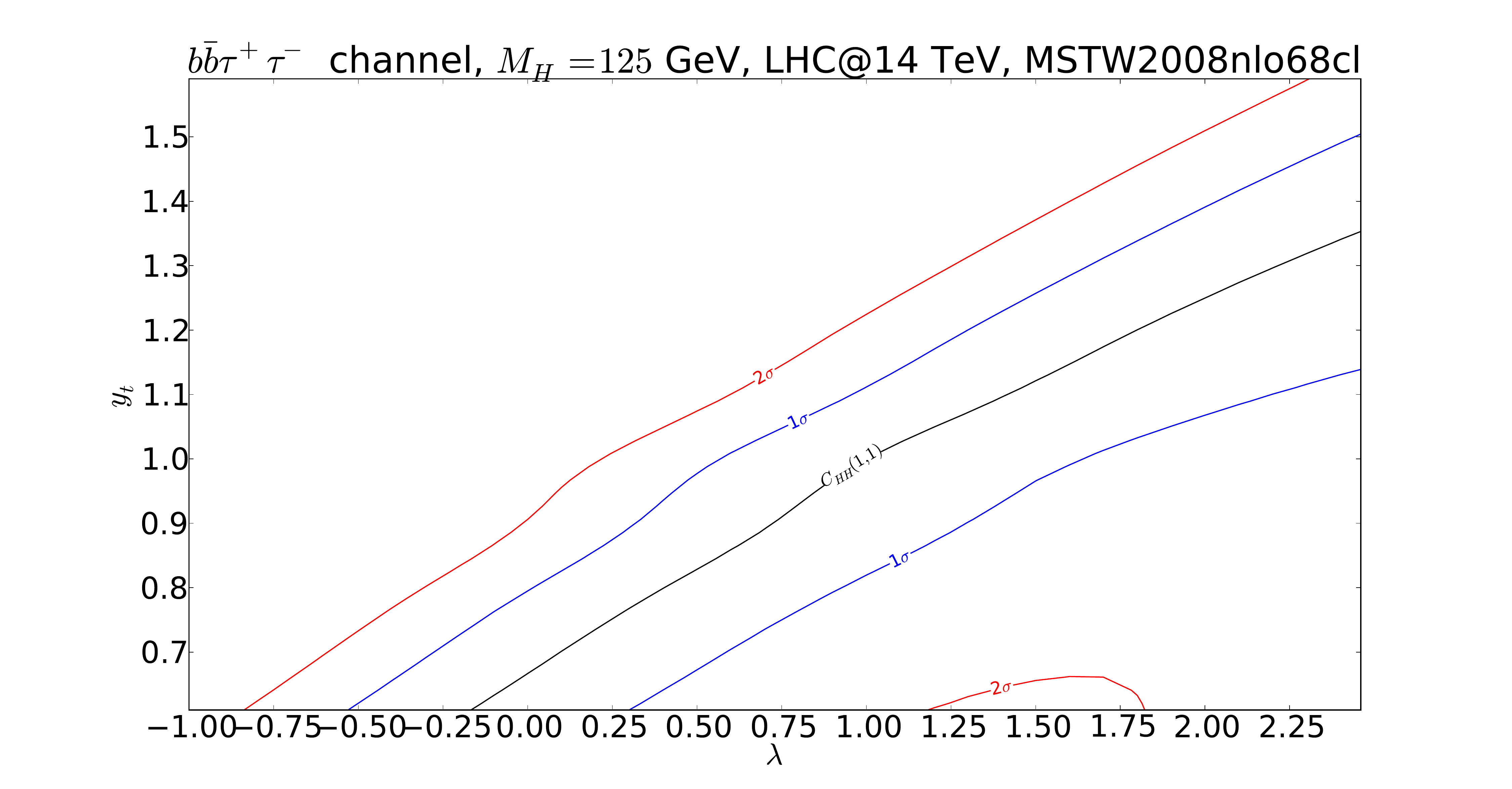}
  \caption{The $1\sigma$ and $2\sigma$ confidence regions in the $y_t-\lambda$ plane at 600~fb$^{-1}$ for the $b\bar{b} \tau^+\tau^-$
    decay mode, derived using $C_{HH}$, within the SM ($\lambda_\mathrm{true} = 1$ and $y_{t,\mathrm{true}} = 1$).}
  \label{fig:ytlambda_tt}
\end{figure}

\begin{figure}[!htb]
  \centering
    \includegraphics[width=0.8\linewidth]{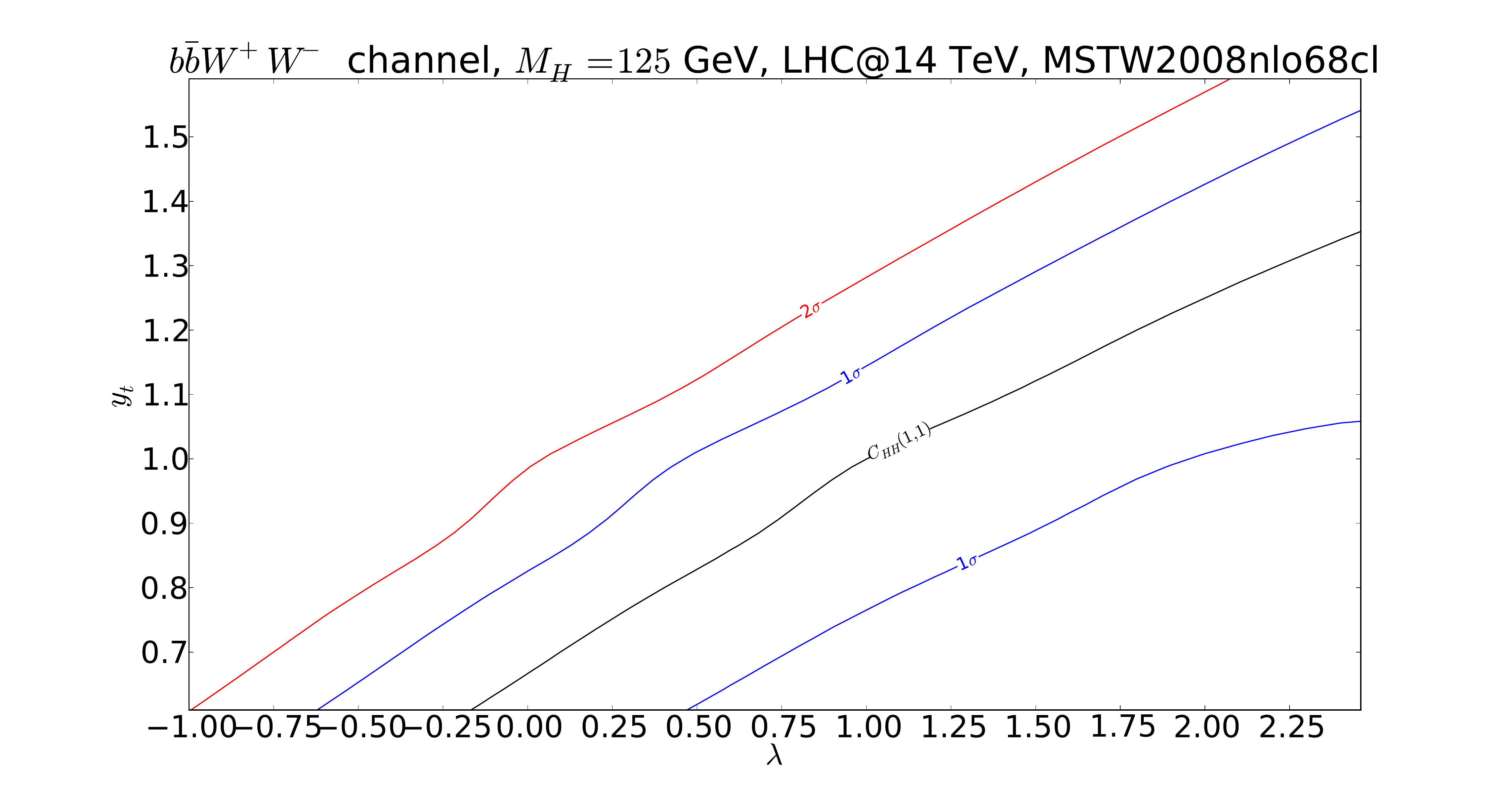}
  \caption{The $1\sigma$ and $2\sigma$ confidence regions in the $y_t-\lambda$ plane at 600~fb$^{-1}$ for the $b\bar{b} W^+W^-$
    decay mode, derived using $C_{HH}$, within the SM
    ($\lambda_\mathrm{true} = 1$ and $y_{t,\mathrm{true}} = 1$). In
    the lower-right corner the exclusion is weak and only the one
    standard deviation curve is shown. }
  \label{fig:ytlambda_ww}
\end{figure}

\begin{figure}[!htb]
  \centering
    \includegraphics[width=0.8\linewidth]{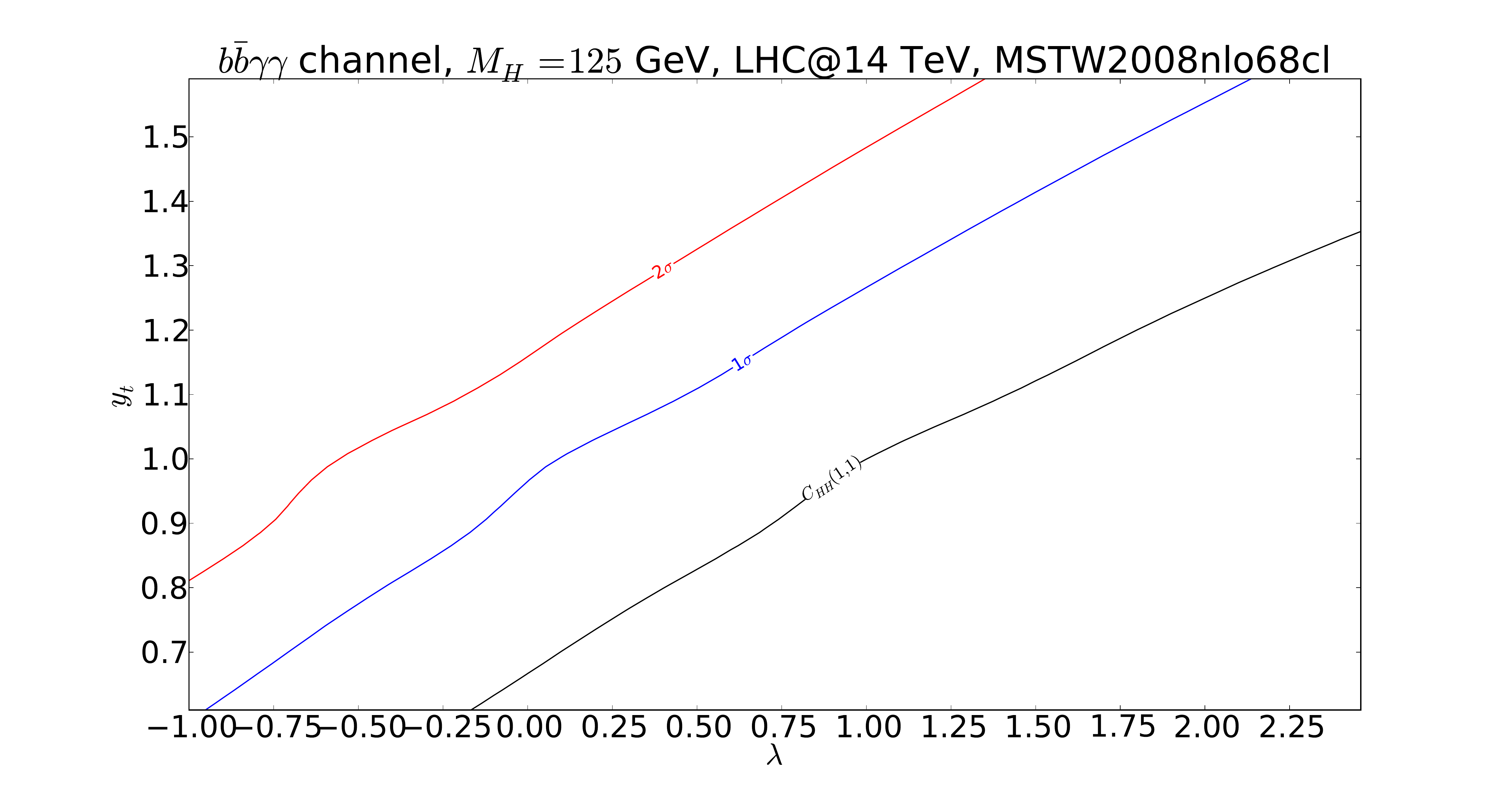}
  \caption{The $1\sigma$ and $2\sigma$ confidence regions in the $y_t-\lambda$ plane at 600~fb$^{-1}$ for the $b\bar{b} \gamma \gamma$
    decay mode, derived using $C_{HH}$, within the SM
    ($\lambda_\mathrm{true} = 1$ and $y_{t,\mathrm{true}} = 1$). In the
    lower-right corner the exclusion is very weak and hence the one and two standard deviation curves are off
    the scale of the figure.}
  \label{fig:ytlambda_gg}
\end{figure}

Since the cross section for
Higgs pair production, as well as the single Higgs cross section, both
depend on the top coupling, a determination of $y_t$ and the
triple coupling, $\lambda$, cannot be done independently through
a measurement of the ratio $C_{HH}$.\footnote{There exist many
  models in which the $Ht\bar{t}$ coupling, $y_t$, can be changed,
  among other effects. See,
  for example, \cite{Agashe:2006wa,
  Casagrande:2008hr, Agashe:2009di, Djouadi:2007fm, Falkowski:2007hz,
  Azatov:2009na, Casagrande:2010si, Patt:2006fw, Schabinger:2005ei}.} The coupling $y_t$ can be deduced by observation of associated production of a single Higgs with top quark
pairs~\cite{Plehn:2009rk} using boosted jet techniques that exploit
the substructure of so-called `fat' jets.\footnote{Note that at the
  LHC no measurements of absolute couplings can be performed. It is
  however possible to make fits to Higgs couplings that are almost
  model-independent using weak theoretical assumptions. For further
  discussion see, for example, Section 2 in
  Ref.~\cite{Peskin:2012we}.} Since the error on a determination of $y_t$ is expected to be
$\mathcal{O}(15\%)$~\cite{Peskin:2012we}, an investigation of the
possible constraints in the $y_t - \lambda$ plane is essential. This
can be done for the Standard Model with the assumption
$\lambda_\mathrm{true} = 1$ and $y_{t,\mathrm{true}} = 1$ in the
simplified model. We can then
calculate the induced error as we have done previously and calculate the $1\sigma$ and $2\sigma$ confidence levels on where
the actual measurement will likely end up in the $y_t - \lambda$ plane. The
results are shown in Figs.~\ref{fig:ytlambda_tt},~\ref{fig:ytlambda_ww} and~\ref{fig:ytlambda_gg} for
$b\bar{b}\tau^+\tau^-$, $b\bar{b}W^+W^-$ and $b\bar{b}\gamma\gamma$
respectively, given an integrated luminosity of 600~fb$^{-1}$. The
figures illustrate an important point: for a model-independent
determination of the Higgs triple self-coupling, a good measurement of $y_t$ is
crucial. If, for example, we consider $y_t$ at the edges of the expected
$\mathcal{O}(15\%)$ error, then $y_t = 0.85$ yields $\lambda \in
(0.2,1.1)$ whereas $y_t = 1.15$ yields $\lambda \in (1.1,\sim 2.4)$,
using the $b\bar{b}\tau^+\tau^-$ channel (Fig.~\ref{fig:ytlambda_tt}), both at $1\sigma$. This is a result of the sensitivity
of the single and double cross sections on $y_t$ (see Eq.~(\ref{eq:sigmaHHfit})). 


\section{Conclusions}\label{sec:conclusions}
We have considered the theoretical error on the ratio of cross
sections of double-to-single Higgs production, $C_{HH}$, at a 14~TeV LHC, including scale
variation and parton density function uncertainties. Under the assumption that the double and single Higgs boson production cross
sections possess a similar form of higher-order corrections, which we motivated in Section~\ref{sec:ratios}, we showed in the same section that the ratio is a more theoretically stable quantity than the cross
section itself. Subsequently, assuming a 5\%
total theoretical error on $C_{HH}$, and using conservative assumptions on the
experimental uncertainties of the quantities involved in measuring the
ratio, we used this ratio to construct possible exclusions in a set of
simplified models, given a true value of the corresponding Higgs self-coupling parameter, at a 14~TeV LHC and
integrated luminosities of 600~fb$^{-1}$ and 3000~fb$^{-1}$. Within the Standard
Model we concluded that it is possible to constrain the trilinear
coupling to be positive, at 95\% confidence level at 600~fb$^{-1}$, only using the discovery of the three viable channels.
We also showed that a naive combination of the `uncertainties' at $1\sigma$ over the three
channels indicates that a measurement of accuracy $\sim +30\%$ and
$\sim -20\%$ is possible simply by using the ratio $C_{HH}$ at 3000~fb$^{-1}$. The
present work outlines the most precise method of determination of the Higgs triple
self-coupling in the SM to date. We have also considered the uncertainty on the top-Higgs coupling and have constructed the possible exclusion
region in the $y_t-\lambda$ plane. Thus, we concluded that an
accurate determination of the $Ht\bar{t}$ coupling, $y_t$, is crucial to the determination of the
Higgs boson triple self-coupling.

It is evident that deviations from expected exclusions within the SM would be an indication of some
inconsistency in these assumptions that would require further
assessment in the form of new physics models. Given the framework that
we have outlined in the present paper, the parameter space relevant to
Higgs pair production can be probed using the ratio $C_{HH}$ in any
BSM theory. Furthermore, it is obvious from the present study, as well as previous ones, that the measurement of the Higgs boson trilinear
self-coupling is a challenging task, and further effort, both on behalf
of theorists and experimentalists, should be made in order to obtain
the best possible measurement during the lifetime of the LHC. 

\section{Acknowledgements} 
We would like to thank Elisabetta Furlan, Thomas Gehrmann, Massimiliano Grazzini, Paolo
Torrielli and Graeme Watt for interesting and useful discussions. We
would also like to thank the Referee for his helpful comments and
suggestions. This work was supported by the Swiss National Science Foundation under contracts
200020-138206 and 200020-141360/1 (AP) and 200020-126632 (FG) and by the Research Executive
Agency of the European Union under the Grant Agreement number
PITN-GA-2010-264564 (LHCPhenoNet). JZ is supported by the ERC Advanced Grant EFT4LHC of the European Research Council, the Cluster of Excellence Precision Physics, Fundamental Interactions and Structure of Matter (PRISMA-EXC 1098).

\bibliography{HHratio.bib}
\bibliographystyle{utphys}

\end{document}